\documentclass[12pt]{iopart}

\usepackage{amsbsy}
\usepackage[english]{babel}
\usepackage{graphicx}
\usepackage{units}

\begin{document}

\title[]{Many-particle effects in adsorbed magnetic atoms with
easy-axis anisotropy: the case of Fe on CuN/Cu(100) surface}

\author{R \v{Z}itko $^{1,2}$, Th Pruschke $^3$}
\address{$^1$ J. Stefan Institute, Jamova 39, SI-1000 Ljubljana, Slovenia}
\address{$^2$ Faculty  of Mathematics and Physics, University of Ljubljana,
Jadranska 19, SI-1000 Ljubljana, Slovenia}
\ead{rok.zitko@ijs.si}
\address{$^3$ Institute for Theoretical Physics, University of G\"ottingen,
Friedrich-Hund-Platz 1, D-37077 G\"ottingen, Germany}

\date{\today}

\begin{abstract}
We study the effects of the exchange interaction between an adsorbed
magnetic atom with easy-axis magnetic anisotropy and the
conduction-band electrons from the substrate. We model the system
using an anisotropic Kondo model and we compute the impurity spectral
function which is related to the differential conductance ($dI/dV$)
spectra measured using a scanning tunneling microscope. To make
contact with the known experimental results for iron atoms on the
CuN/Cu(100) surface [Hirjibehedin et al., Science {\bf 317}, 1199
(2007)], we calculated the spectral functions in the presence of an
external magnetic field of varying strength applied along all three
spatial directions. It is possible to establish an upper bound on the
coupling constant $J$: in the range of the magnetic fields for which
the experimental results are currently known (up to $\unit[7]{T}$),
the low-energy features in the calculated spectra agree well with the
measured $dI/dV$ spectra if the exchange coupling constant $J$ is at
most half as large as that for cobalt atoms on the same surface. We
show that for even higher magnetic field (between 8 and $\unit[9]{T}$)
applied along the ``hollow direction'', the impurity energy states
cross, giving rise to a Kondo effect which takes the form of a
zero-bias resonance. The coupling strength $J$ could be determined
experimentally by performing tunneling spectroscopy in this range of
magnetic fields. On the technical side, the paper introduces an
approach for calculating the expectation values of global spin
operators and all components of the impurity magnetic susceptibility
tensor (including the out-of-diagonal ones) in numerical
renormalization group (NRG) calculations with no spin symmetry.  An
appendix contains a density-functional-theory (DFT) study of the Co
and Fe adsorbates on CuN/Cu(100) surface: we compare magnetic moments,
as well as orbital energies, occupancies, centers, and spreads by
calculating the maximally localized Wannier orbitals of the
adsorbates.
\end{abstract}


\maketitle

\newcommand{\vc}[1]{{\boldsymbol{#1}}}
\newcommand{\ket}[1]{|#1\rangle}
\newcommand{\bra}[1]{\langle #1|}
\newcommand{\braket}[1]{\langle #1 \rangle}
\newcommand{\expv}[1]{\braket{#1}}
\renewcommand{\Im}{\mathrm{Im}}
\renewcommand{\Re}{\mathrm{Re}}
\newcommand{\dr}{\mathrm{d}}
\newcommand{\correl}[1]{\langle\langle #1 \rangle\rangle}
\newcommand{\TKO}{T_K^{(0)}}
\newcommand{\TKt}{T_K^{(2)}}

\newcommand{\figw}{7cm}
\newcommand{\dIdV}{\mathrm{d}I/\mathrm{d}V}

\bibliographystyle{unsrt}

\section{Introduction}

Recent advances in the experimental techniques, such as
low-temperature scanning tunneling microscopy (STM) and the X-ray
magnetic circular dichroism (XMCD), made it possible to study magnetic
properties of single adsorbed atoms on various surfaces
\cite{gambardella2003, heinrich2004, hirjibehedin2006, meier2008,
balashov2009, ternes2009, brune2009}. Particularly interesting are
magnetic impurities on noble metal surfaces where many-particle
effects such as the Kondo effect play an important role \cite{li1998,
madhavan1998}. Due to the reduced symmetry in the surface region, such
magnetic adsorbates are strongly anisotropic \cite{gambardella2003,
hirjibehedin2007}. A simple parametrization of the leading magnetic
anisotropy terms takes the form of
\begin{equation}
\label{eq1}
H_\mathrm{aniso} = D S_z^2 + E(S_x^2-S_y^2),
\end{equation}
where the $S_x$, $S_y$, and $S_z$ are the cartesian components of the
quantum-mechanical spin-$S$ operator, while the parameters $D$ and
$E$ are known as the longitudinal and transverse magnetic anisotropy,
respectively. If $D<0$ (the ``easy-axis'' case), the spin tends to
maximize the absolute value of the $z$ component, thus the $S_z=\pm S$
states dominate in the ground state doublet, which may be split for
integer $S$ even in the absence of an external magnetic field due to
the transverse magnetic anisotropy $E$. If $D>0$ (the ``easy-plane''
or ``hard-axis'' case), the ground-state is a doublet consisting
mostly of the $S_z=\pm 1/2$ states for half-integer spin, or a singlet
consisting mostly of the $S_z=0$ state for integer spin. Similar
models are also used to describe molecular magnets \cite{romeike2006,
romeike2006b, leuenberger2006, wegewijs2007, gonzalez2008}. 

When adsorbed on metallic surfaces, the coupling of the spin to the
conduction-band electrons can introduce new phenomena, the Kondo
effect being the most interesting due to its unique signature expected
in the tunneling spectra. For half-integer spins with easy-plane
anisotropy, the level degeneracy in the ground state actually permits
the occurrence of the Kondo effect at zero magnetic field, as indeed
observed experimentally in the system of cobalt atoms adsorbed on the
CuN ultra-thin layers deposited on the Cu(100) substrate surface
\cite{otte2008}. For magnetic atoms with easy-axis anisotropy,
however, the Kondo effect has not yet been observed. Its absence is
either due to the zero-field level splitting caused by the transverse
magnetic anisotropy $E$, which in the case of integer $S$ plays the
role of an effective magnetic field which suppresses the Kondo effect,
or (for half-integer $S$ or for $E \ll |D|$) due to the fact that the
spin-flip scattering events of the conduction-band electrons can only
change the impurity spin component by 1. Thus the two impurity
ground-state levels only couple via higher-order processes and the
Kondo screening is strongly suppressed, i.e., the Kondo scale is
extremely small \cite{romeike2006b}. Nevertheless, interesting
many-particle effects in fact do occur in the case of easy-axis
anisotropy, thus we study this class of the problems in the following
parts of the paper, focusing in particular on the case of spin 2 as
appropriate for iron atoms adsorbed on the CuN/Cu(100) substrate
\cite{hirjibehedin2007}.

The absence of the Kondo effect at zero magnetic field does not imply,
that the exchange coupling of an easy-axis magnetic impurity to the
substrate is of no consequence. In this work we show that the value of
the exchange coupling constant $J$, defined through the Kondo coupling
Hamiltonian
\begin{equation}
\label{eq2}
H_\mathrm{K} = J \vc{s} \cdot \vc{S},
\end{equation}
where $\vc{s}$ is the spin-density of the conduction-band electrons at
the position of the impurity, affects the impurity spectral function
in a measurable way. The most drastic effect is the occurrence of a
Kondo effect induced by a magnetic field, similar to the
singlet-triplet Kondo effect in quantum dot systems \cite{sasaki2000,
pustilnik2000, izumida2001, pustilnik2001st, pustilnik2003,
izumida2001, hofstetter2004}: for some high enough external magnetic
field applied along the $x$-axis, the impurity level crossing leads to
a situation where the conditions for the occurrence of a Kondo effect
are met. We will show that the presently available experimental
results (measured up to $\unit[7]{T}$, Ref.~\cite{hirjibehedin2007})
just stop short of the region where this Kondo resonance should be
observed (between $\unit[8]{T}$ and $\unit[9]{T}$). For this reason,
the presently known experimental results can only be used to establish
an upper bound on the value of $J$, but by extending the magnetic
field range into the $\unit[9]{T}$ range, the Kondo resonance could be
probed and the exchange coupling $J$ might be determined even
quantitatively.

The paper is structured as follows. In Sec.~\ref{sec2} we provide the
full definition of the model under study and describe the numerical
technique which is used to compute the impurity spectral function. We
also comment on the relation between the impurity spectral function
and the experimentally measured $dI/dV$ spectrum. In Sec.~\ref{sec3}
we discuss the impurity level structure and the predicted $dI/dV$
spectra in the limit of a decoupled impurity, $J \to 0$. Such spectra
do not encompass any many-particle effects; however, they account
accurately for the high-energy spin excitations which are observed
experimentally. In Sec.~\ref{sec4} the results of the numerical
calculations for the full many-particle problem are presented and
discussed in relation with the experimental results.  \ref{appa}
presents a comparative density-functional-theory (DFT) study of Co and
Fe adatoms on the CuN/Cu(100) surface, while \ref{appb} contains the
description of a method for computing the thermodynamic impurity spin
susceptibility in problems having no symmetries in the spin space with
the numerical renormalization group (NRG) method.

\section{Model and method}
\label{sec2}

We model the system using the Hamiltonian $H = H_\mathrm{band} +
H_\mathrm{K} + H_\mathrm{imp}$ where
\begin{eqnarray}
H_\mathrm{band} &= \sum_{k\sigma} \epsilon_k c^\dag_{k\sigma} c_{k\sigma}, \\
H_\mathrm{imp} &= g\mu_B \vc{B} \cdot \vc{S} + H_\mathrm{aniso}.
\end{eqnarray}
The anisotropy term $H_\mathrm{aniso}$ is given by Eq.~\eref{eq1}, and
the coupling term $H_\mathrm{K}$ by Eq.~\eref{eq2}. The operator
$c_{k\sigma}$ corresponds to a conduction-band electron with momentum
$k$, spin $\sigma \in \{\uparrow,\downarrow\}$, and energy
$\epsilon_k$. In terms of these operators, the spin density of the
conduction-band electrons at the position of the impurity (assumed to
be at the origin of the coordinate system) is given by $\vc{s} = (1/2)
f^\dag_{0,\alpha} \boldsymbol{\sigma}_{\alpha\beta} f_{0,\beta}$,
where
\begin{equation}
f_{0\sigma} = \frac{1}{\sqrt{N}} \sum_{k} c_{k\sigma},
\end{equation}
$N$ is the number of the conduction-band states, and
$\boldsymbol{\sigma}_{\alpha\beta}$ is the vector of Pauli matrices.
Finally, $g$ is the gyromagnetic factor (assumed to be isotropic), and
$\mu_B \approx 0.0579 \unit{meV/T}$ is the Bohr magneton. For an iron
adatom on the CuN/Cu(100) surface, the following parameters have been
established: spin $S=2$, $g=2.11 \pm 0.05$, $D=
\unit[-1.55\pm0.01]{meV}$, $E=\unit[0.31\pm0.01]{meV}$
(Ref.~\cite{hirjibehedin2007}). For a complete description one should,
furthermore, know the density of states (DOS) of the conduction band,
$\rho(\omega)=(1/N)\sum \delta(\omega-\epsilon_k)$, and the value of
the Kondo coupling $J$. In the interesting energy range (on the order
of $\unit[10]{meV}$ around the Fermi level) the energy dependence of
the DOS may be neglected and we can use a constant value
$\rho=\rho(\omega=0)$. The low-energy behavior of the problem then
depends solely on the dimensionless combination $\rho J$. We note that
the model does not include any orbital moment degrees of freedom. This
approximation is based on the fact that the orbital moment of an
impurity adsorbed on the surface is typically strongly quenched due to
the surface field effects (for the particular case of Fe and Co
adsorbates on the surface of CuN/Cu(100), the total suppression of
orbital degeneracy follows from the DFT results tabulated in
\ref{appa}).

We solve the problem using the numerical renormalization group
\cite{wilson1975, krishna1980a, bulla2008} (for the details on the
application of the NRG to the problems of this class see also
Refs.~\cite{aniso, aniso2}). In this work, the focus of our interest
will be the impurity spectral function defined as \cite{costi2000}
\begin{equation}
A_\sigma(\omega) = -\frac{1}{\pi} \Im T_\sigma(\omega+i\delta),
\end{equation}
where $T_\sigma(\omega)$ is the T-matrix for the scattering of electrons of
spin $\sigma$ on the spin-$S$ impurity, given by
\begin{equation}
T_\sigma(\omega) = \frac{J}{2} \expv{S_z} +
\correl{O_\sigma;O_\sigma^\dag}_\omega,
\end{equation}
with the operators
\begin{eqnarray}
O_\uparrow   &= \frac{J}{2} \left( f_{0\downarrow} S^{-}
+ f_{0\uparrow} S_z \right), \\
O_\downarrow &= \frac{J}{2} \left( f_{0\uparrow} S^{+}
- f_{0\downarrow} S_z \right).
\end{eqnarray}
Here $S^\pm=S_x \pm i S_y$. The notation $\correl{A;B}_\omega$ denotes
the Laplace transform of the correlator $\correl{A;B}_t
=-i\theta(t)\langle \{A(t),B(0)\}_+ \rangle$ between two (fermionic)
operators $A$ and $B$, i.e., $\correl{A;B}_\omega=\int_0^\infty
e^{i\omega t} \correl{A;B}_t \mathrm{d}t$ with $\mathrm{Im}\,\omega>0$;
$\{A,B\}_+=AB+BA$ is the anti-commutator.

The T-matrix itself is defined through
\begin{equation}
G_{kk'\sigma}(\omega) = \delta_{kk'} G_{kk'\sigma}^{0}(\omega)
+G^0_{kk\sigma}(\omega) \frac{1}{N} T_\sigma(\omega) G^0_{k'k'\sigma}(\omega),
\end{equation}
where $G_{kk\sigma}(\omega)= \correl{c_{k\sigma} ; c_{k\sigma}^\dag}$ is the
conduction-band electron Green's function and $G^0_{kk\sigma}$ its
unperturbed counterpart. In the STM experiments, the spin components are not
resolved, thus we will mostly plot the spin-averaged spectral functions,
$A(\omega) = [A_\uparrow(\omega)+A_\downarrow(\omega)]/2$.

\section{Decoupled impurity limit}
\label{sec3}

\begin{figure}[htbp]
\centering
\includegraphics[clip,width=14cm]{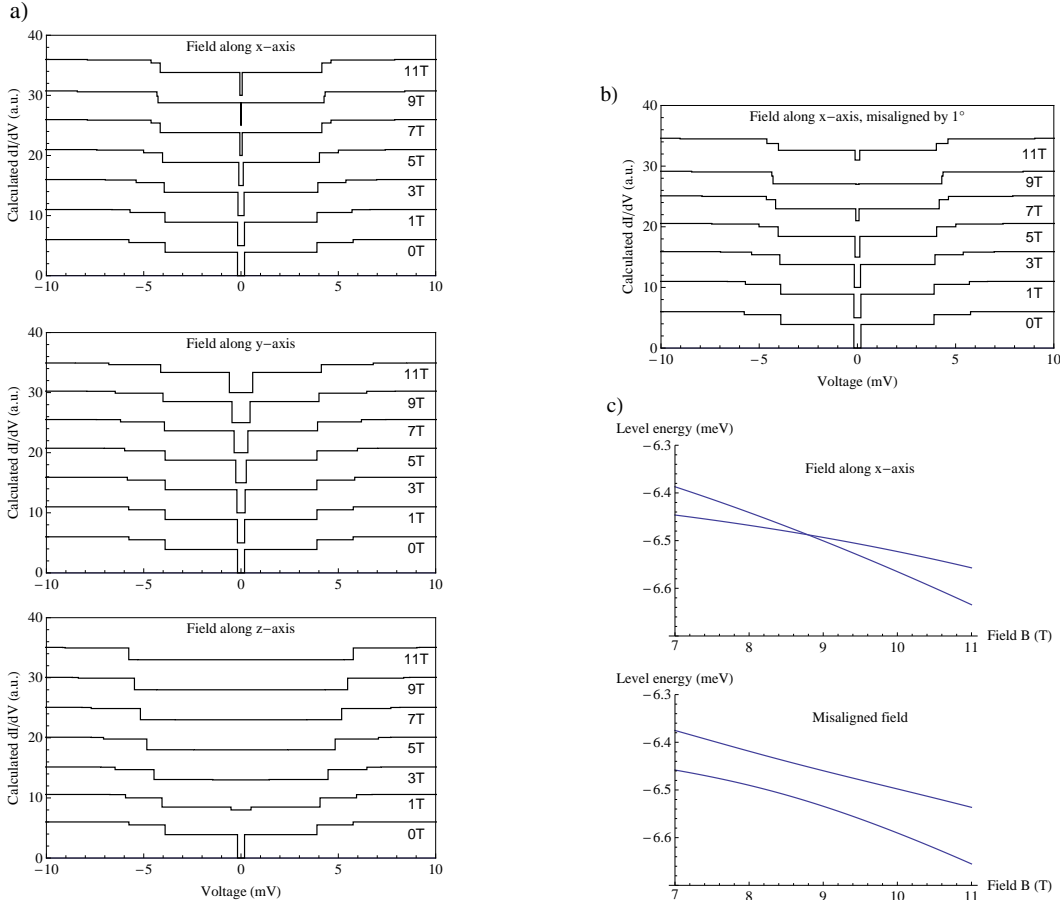}
\caption{a) Simulated IETS spectra. No thermal broadening has been applied.
b) Simulated IETS spectra for a field with a direction almost along the
$x$-axis with a small component along the $z$-axis (misalignment by
$\unit[1]{{}^\circ}$. c) Energy-level diagrams for a field along the
$x$-axis (top) and for a field misaligned by $\unit[1]{{}^\circ}$ (bottom).}
\label{fig1}
\end{figure}

The spin excitation spectrum of a magnetic atom adsorbed on a thin
insulating layer may for the most part be accounted for by fully
neglecting its exchange coupling to the substrate electrons. According
to this picture, the tunneling differential conductance will increase
stepwise when the bias voltage is increased beyond the values
$(E_i-E_{gs})/e$, where $E_i$ are the energies of the excited states
and $E_{gs}$ is the ground-state energy \cite{heinrich2004,
hirjibehedin2006, hirjibehedin2007}. At each of these voltages, an
additional inelastic scattering (spin-flip) channel opens, typically
resulting in an increased conductance \cite{jaklevic1966, lambe1968,
persson1987, ho2002, anderson1966, appelbaum1966, scalapino1967,
appelbaum1967, maltseva2009}. In fact, the experiments have shown that
the conductance step heights are described to a good approximation by
\cite{heinrich2004}
\begin{equation}
|\bra{i} S_x \ket{gs}|^2
+
|\bra{i} S_y \ket{gs}|^2
+
|\bra{i} S_z \ket{gs}|^2,
\end{equation}
i.e., by the transition matrix elements for the spin operator. This
empirical observation has recently received theoretical support
\cite{lorente2009, fernandezrossier2009}. The expression suggests that
the impurity is well thermalized with the substrate: the impurity
relaxes to the ground-state on a time-scale which is short compared
with the mean time between the tunneling events. Using this simple
approach, one can compute $dI/dV$ spectra which compare very well with
the experimental results (as long as the many-particle effects are not
important) if thermal broadening is taken into account. Here we show
the results obtained by this procedure without adding any artificial
thermal smearing, see Fig.~\ref{fig1}a. The plots should be compared
with Fig.~2A,B,F and Fig.~S1,A,C in Ref.~\cite{hirjibehedin2007}.
Presenting the results in the zero-temperature limit uncovers more
details, in particular the crossing of the the levels in the case of a
field applied along the $x$-axis (this corresponds to the ``hollow
axis'' in Ref.~\cite{hirjibehedin2007}).

As is well known, the experimental results are in very good agreement
with the simulated $dI/dV$ spectra \cite{hirjibehedin2007}. One
comment, however, is in order. In experiments, the step height of the
first step in the dI/dV spectra is observed to decrease as the field
strength along the $x$-axis is increased. This can be explained, on
one hand, by the thermal broadening of the spectra which leads to
reduced heights of spectral features of the width smaller than the
width of the broadening kernel. Another explanation is based on a
possibility of a slight misalignment of the magnetic field from the
direction of the actual $x$-axis of the magnetic atom. A simple
calculation (Fig.~\ref{fig1}b) shows that even a small misalignment of
$\unit[1]{{}^\circ}$ (which is quite likely to occur) can suppress the
step height of the first excitation (at $T=0$) without significantly
modifying those at the other magnetic excitation energies.

We now comment on the important observation that the energy difference
between the two lowest levels decreases as the field strength along the
$x$-axis is increased, see Fig.~\ref{fig1}c, top. A simple calculation
neglecting any possible many-particle effects indicates that the levels
become degenerate at
\begin{equation}
\label{eqbx}
B_x = \frac{1}{g\mu_B} \sqrt{2E(E-D)} \approx \unit[8.8]{T}.
\end{equation}
For this field strength, a magnetic-field-induced Kondo effect can
occur if the Kondo exchange coupling is antiferromagnetic, $J>0$. We
note that the aforementioned misalignment of the field will also tend
to suppress the Kondo effect at $B_x \approx \unit[8.8]{T}$, since a
component of the field along the $z$-axis will mix the levels, thereby
leading to an avoided crossing of the levels (Fig.~\ref{fig1}c,
bottom). In experiments seeking to detect the field-induced Kondo
effect, it will be thus essential to not only fine-tune the magnitude
of the magnetic field, but also its exact direction.

\section{Impurity spectral functions}
\label{sec4}

\begin{figure}[htbp]
\centering 
\includegraphics[clip,width=8cm]{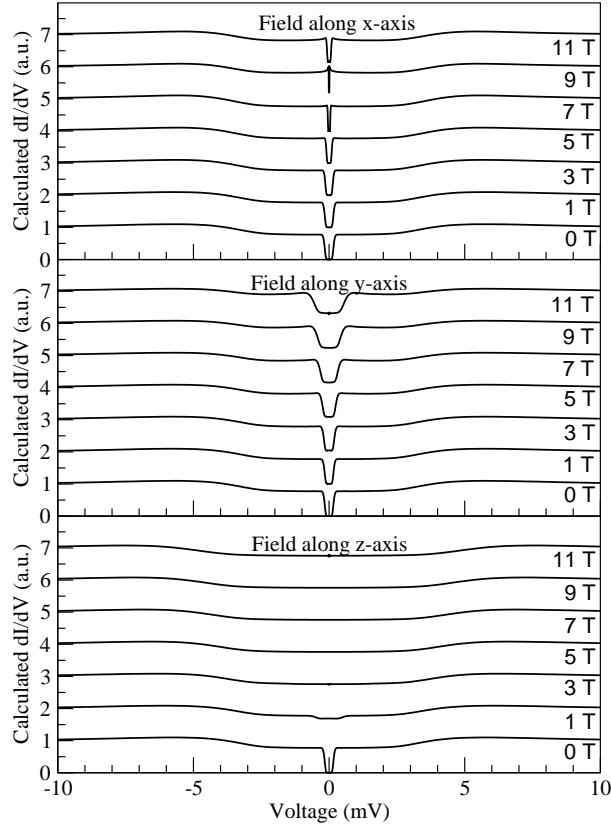}
\caption{Impurity spectral functions $A(\omega)$ computed using the
NRG. The horizontal energy axis has been transformed to show the
equivalent bias voltages, while the vertical axis corresponds to the
$dI/dV$ differential conductance in the simple approximation where the
LDOS is taken to be proportional to the spectral function. Here $\rho
J=0.05$.} \label{fig2}
\end{figure}

In Fig.~\ref{fig2} we plot the impurity spectral functions $A(\omega)$
computed using the numerical renormalization group (NRG). The
calculation is performed for zero temperature, thus the widths of
spectral features are due to intrinsic effects (the NRG spectral
function overbroadening artifacts are small in these calculations; for
discussion see Ref.~\cite{resolution}). Compared with the simulated
$dI/dV$ curves, all spectral features except for those near the Fermi
level are almost completely smeared out. This is at odds with the
experimental results, where the steps are observed to be sharp apart
from the thermal broadening. While the calculated spectral function
does encompass some inelastic scattering effects (see Fig.~3 in
Ref.~\cite{aniso2}), the tunneling current includes additional
inelastic scattering contributions which are beyond the scope of our
calculations that are constrained to the case of the magnetic impurity
being in thermal equilibrium with the substrate electron reservoir at
all times and do not include any additional interactions between the
tunneling electron and the impurity local moment. Nevertheless, the
results obtained from a NRG calculation provide a reliable description
of the behavior of the system in the immediate vicinity of the Fermi
level, which is probed by tunneling electrons of sufficiently
low-energy so that the system is not driven significantly out of the
equilibrium.

\begin{figure}[htbp]
\centering
\includegraphics[clip,width=8cm]{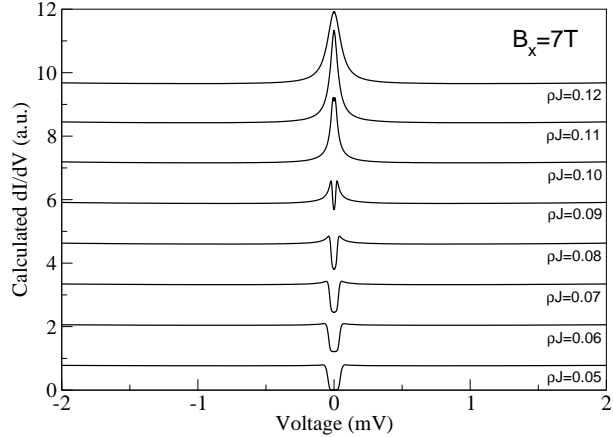}
\caption{Impurity spectral functions for $\unit[7]{T}$
magnetic field applied along the $x$-axis for a range of $\rho J$.
}
\label{fig3}
\end{figure}

The agreement between the low-energy features (interval
$[-\unit[1]{mV}:+\unit[1]{mV}]$) in the calculated and measured
spectra \cite{hirjibehedin2007} in the 0 to \unit[7]{T} range is very
good if thermal broadening is taken into account. The calculation was
performed for $\rho J=0.05$. This value is half as large as the
exchange coupling for Co impurities on the same surface, $\rho J=0.1$,
which has been established in Refs.~\cite{aniso, aniso2}) by
determining the value of $\rho J$ that allows to reproduce the
experimental differential conductance spectra for the Co/CuN/Cu(100)
adsorbate system. Due to similarities between Co and Fe atoms the
exchange coupling constants should not be too different, but very
likely of the same order of the magnitude and of the same sign (Co and
Fe are neighbours in the periodic system; see \ref{appa} for
a more rigorous comparative study using the density functional
theory).  In Fig.~\ref{fig3}a we compare the low-energy part of the
spectral function for a fixed magnetic field of \unit[7]{T} in the
$x$-axis direction for a range of values of $J$. For large $J$, the
dip in the spectral density turns into a (Kondo) resonance which
becomes fully developed for $\rho J \geq 0.1$. For $\rho J$ in the
interval 0.06 to 0.1, we find an intermediate structure: a dip with
protrusions at the step edges. At high enough temperature, the
protrusions will be washed out and a single antiresonance would be
detected. Taking all these elements into account, the estimate of the
upper bound $\rho J=0.05 \pm 0.02$ therefore seems reasonable.

The magnetic field at which the Kondo effect occurs for integer spin
Fe is not given exactly by Eq.~\eref{eqbx}, but it rather depends on
the value of $J$, since the different impurity levels are renormalized
differently due to the anisotropy. In the $J \to 0$ limit, the Kondo
effect occurs at $B_x \approx \unit[8.8]{T}$, while for finite $J$ the
magnetic field where the Kondo peak may be observed is reduced. In
addition, the parameter $J$ controls the width of the Kondo resonance
(and, of course, the Kondo temperature itself), thus for smaller $J$
the interval of the magnetic fields where the Kondo effect may be
observed is accordingly narrower and the Kondo scale lower.

\begin{figure}[htbp]
\centering
\includegraphics[clip,width=8cm]{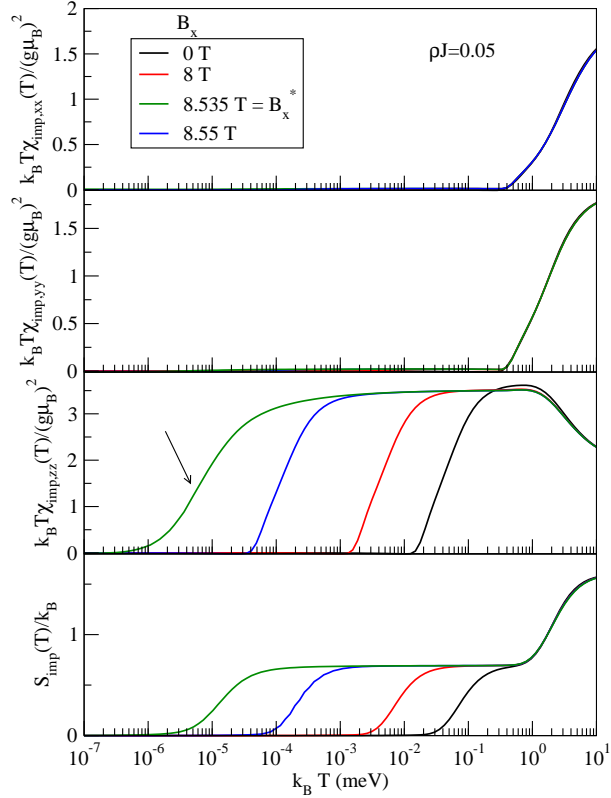}
\caption{Thermodynamic properties for a range of magnetic fields along 
the $x$-axis. We plot the impurity contribution to the magnetic susceptibility
in various directions, $\chi_{\mathrm{imp},\alpha\alpha}$, where $\alpha
\in \{x,y,z\}$ is a direction in space, and the impurity contribution
to the entropy, $S_\mathrm{imp}$.
The Kondo temperature is estimated from the data for
$S_\mathrm{imp}$ to be $T_K \sim \unit[10^{-5}]{meV} \sim \unit[0.1]{mK}$. 
 The arrow points to the curve with
exhibits Kondo-like temperature dependence.}
\label{td}
\end{figure}

The magnetic-field-induced Kondo effect exhibits a number of rather
peculiar features. We first note that it is very difficult to observe
the fully developed Kondo resonance unless $J$ is rather large. For
the estimated exchange coupling of $\rho J=0.05$, where the Kondo
effect occurs for $B_x^* = \unit[8.535]{T}$, the Kondo temperature is
actually extremely low, on the order of $\unit[0.1]{mK}$, see
Fig.~\ref{td}. Even at dilution refrigerator temperatures, it thus
appears unlikely that a fully developed Kondo resonance will be
observed. This does not, however, preclude the experimental detection
of the Kondo physics at play in this system, since the evolution of
the tunneling spectrum as the magnetic field is tuned should exhibit 
a characteristic merging of the two ``half-resonances''.

The results for the thermodynamic properties also reveal 
that the $zz$ component of the magnetic susceptibility multiplied by
the temperature [i.e., the effective moment $T\chi_{zz}(T)$] exhibits
a plateau at very high value of 3.5 (in suitable units), which is
rather near the limiting value for a decoupled $S_z=\pm 2$ doublet,
which is 4. This plateau corresponds to a $\ln 2$ plateau in the
impurity entropy, i.e., to an effective two-state system. On the scale
of $T_K$, the moment is then screened. Note that the temperature
dependence of $T \chi_{zz}(T)$ is different (Kondo-like, see the arrow
in Fig.~\ref{td}) for $B_x=B_x^*$ as compared with those for other
values of the magnetic field, which merely follow the $\exp(-\Delta
E/k_B T)$ rule, where $\Delta E$ is the difference between the
two lowest impurity level energies.

\begin{figure}[htbp]
\centering 
\includegraphics[clip,width=10cm]{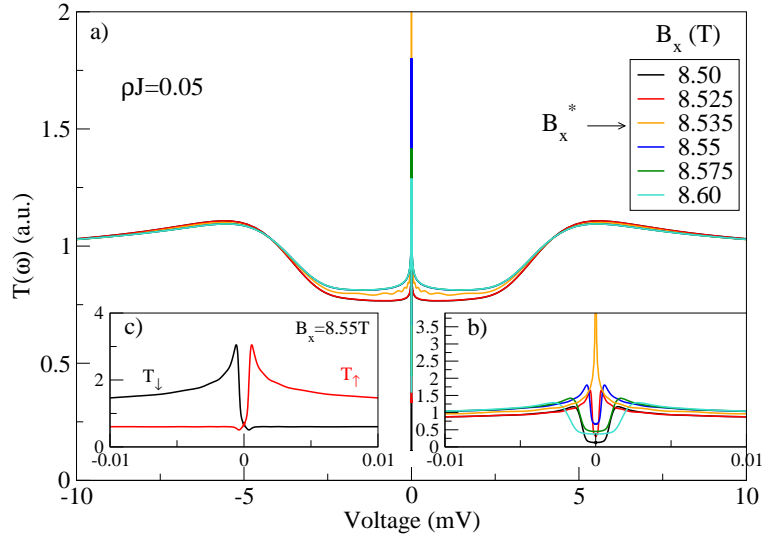}
\caption{Impurity spectral functions in the vicinity of the
magnetic-field-induced Kondo effect at $B_x=B_x^*$. a) Overview plot,
b) close-up on the low-voltage region. The jaggedness of the curve at
the transition point is due to increased numeric artifacts related to
the ground-state crossing. c) Spin-resolved spectral function away
from the transition point. Note the rather sharp features in the
field-split Kondo resonance.} \label{disc}
\end{figure}

It is equally interesting to note that the spectral functions change
discontinuously as we cross the transition value $B_x^*$, see
Fig.~\ref{disc}a; this corresponds to the system being in different
ground states on either side of the transition. In the presence of a
small magnetic field component along the $z$-axis, the level-crossing
is replaced by a rapid cross-over. Any such rapid change of the
spectral function over an extended energy interval would be an
indicator of the proximity to the Kondo regime, even at higher
temperatures.

\begin{figure}[htbp]
\centering
\includegraphics[clip,width=12cm]{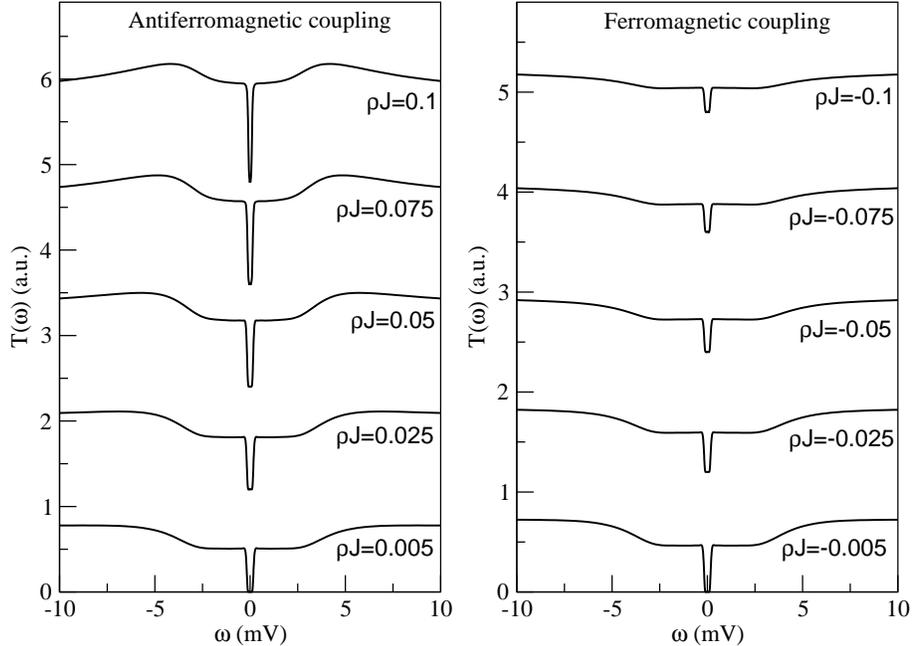}
\caption{Impurity spectral functions at zero field for different
values of exchange coupling $J$.}
\label{fig4}
\end{figure}

The magnetic-field-induced Kondo effect is not the only observable
consequence of the exchange coupling with the substrate. The value (as
well as the sign) of $J$ is actually reflected in the spectral
function for any value of the magnetic field. In Fig.~\ref{fig4} we
compare, for example, the impurity spectral function at zero field for
a range of coupling constants. The exchange coupling strongly affects
the features around the Fermi level, but also those at higher
energies, although to a lesser degree.

The results which we have established for the $S=2$ Kondo model also
apply more generally for other integer spins: at some $B_x^*$ there
will be level crossing which can induce a Kondo resonance at the Fermi
level, but no such crossing occurs for the magnetic field along the
$y$ and $z$ axis. The same is true for half-integer spin models, but
in this case there is, in addition, a two-fold level degeneracy also
for zero magnetic field, thus for half-integer-spin magnetic
impurities one could observe both the zero-field Kondo effect and the
magnetic-field-induced Kondo effect.

\section{Conclusion}

We have studied thermodynamic and spectral properties of the Kondo
model with an easy-axis anisotropy. By comparing the calculated
spectral functions with the measured ones in the magnetic field range
up to $\unit[7]{T}$ we have estimated the upper bound for the exchange
coupling of a Fe adatom on CuN/Cu(100) surface with the bulk
conduction-band electrons to be $\rho J = 0.05 \pm 0.02$. By extending
the range of the magnetic fields in the experiments up to
$\unit[9]{T}$, it should be possible to obtain an improved upper bound
on $J$ or perhaps even observe a magnetic-field-induced Kondo
resonance if $J$ is high enough. We note here that, strictly speaking,
the quantity $J$ scales under the renormalization-group flow, thus its
actual value depends on the choice of the bandwidth and the
corresponding density of states $\rho$; a more rigorous treatment of
the problem would take into account the full density-of-states
dependence of the substrate and, furthermore, the momentum dependence
of the Kondo coupling $J(\vc{k},\vc{k}')$, see for example
Ref~\cite{hewson}. Nevertheless, a simple characterization in terms of
the dimensionless combination $\rho J$ is sufficient to compare the
over-all effect of the Kondo screening for different adsorbed atoms.

For quantitative estimates, it will be important also to reduce the
temperature at which the experiments are performed as far as possible.
Irrespective if the Kondo resonance around $\unit[8.5]{T}$ will
eventually be observed or not, the real motivation for performing such
experiments actually comes from the need to estimate the exchange
coupling to the bulk electrons, since this parameter also determines
the spin relaxation time; thus it plays an important role in the
possible applications of high-spin states of magnetic impurities for
storing and processing information.

\appendix

\section{DFT study of Co and Fe adatoms on CuN/Cu(100)}
\label{appa}

\begin{table}[htb]
\caption{
\label{table0}
Properties of the Co and Fe adatom on CuN/Cu(100) surface.}
\begin{indented}
\item[]\begin{tabular}{@{}lllll@{}}
\br
System                     
                           & \multicolumn{1}{c}{Moment ($\mu_B$/cell)}
                           & \multicolumn{1}{c}{$n_d$}
                           & \multicolumn{1}{c}{$n_{d\uparrow}-n_{d\downarrow}$}
                           & \multicolumn{1}{c}{$n_s$} \\
\mr
Co/Cu$_2$N/Cu(100)         & 2.88 & 7.56 & 2.10 & 0.64 \\
Fe/Cu$_2$N/Cu(100)         & 4.12 & 6.61 & 3.09 & 0.63 \\
\br
\end{tabular}
\end{indented}
\end{table}

\begin{table}[htb]
\caption{\label{table1co} The properties of the MLWFs of Co on
CuN/Cu(100). We tabulate orbital energy $\epsilon$, occupancy
$n$, orbital spread $s$, and orbital center along the $z$-axis. The
orbital labels correspond to the initial projections used in the
calculation of the corresponding maximally localized Wannier orbitals,
although the orbitals are significantly modified from the free-atom
orbitals due to the strong hybridization with the neighboring N atoms
and the Cu atom just below the adsorbate. The position of the Co
adatom in the $z$-direction is $z(\mathrm{Co})=$\unit[-0.148]{\AA};
the origin for the $z$-axis corresponds to the top-most layer of Cu
atoms prior to relaxation. }
\lineup
\begin{indented}
\item[]\begin{tabular}{@{}lllllllll@{}}
\br
{\bf Co}               & \multicolumn{4}{c}{Minority spin $\downarrow$}
               & \multicolumn{4}{c}{Majority spin $\uparrow$} \\ \ns
 & \crule{4} & \crule{4} \\
Orbital        & \multicolumn{1}{c}{$\epsilon$ (eV)}
               & \multicolumn{1}{c}{$n$}
               & \multicolumn{1}{c}{$s$ (\AA$^2$)}
               & \multicolumn{1}{c}{$z$ (\AA)}
               & \multicolumn{1}{c}{$\epsilon$ (eV)}
               & \multicolumn{1}{c}{$n$}
               & \multicolumn{1}{c}{$s$ (\AA$^2$)}
               & \multicolumn{1}{c}{$z$ (\AA)} \\
\mr
$s$            & \01.70   & 0.27  &  2.36 & -3.06   &  \01.32  & 0.37  &  2.07 & -3.03 \\
$d_{3z^2-r^2}$ & -1.01  & 0.88  &  0.73 & -2.01   & -2.41  & 0.97  &  0.52 & -2.06 \\
$d_{zx}$       & -0.07  & 0.28  &  0.94 & -2.09   & -2.83  & 1.00  &  0.42 & -2.15 \\
$d_{zy}$       & -0.50  & 0.41  &  0.50 & -2.17   & -2.98  & 0.97  &  0.44 & -2.17 \\
$d_{x^2-y^2}$  & -1.04  & 0.80  &  1.16 & -2.03   & -2.30  & 0.90  &  0.77 & -2.08 \\
$d_{xy}$       & -0.34  & 0.36  &  0.65 & -2.12   & -2.90  & 0.99  &  0.46 & -2.13 \\
$d$ total      &        & 2.73  &          &    &           & 4.83  & \\
\br
\end{tabular}
\end{indented}
\end{table}

\begin{table}[htb]
\caption{\label{table1fe} The properties of the MLWFs of Fe on
CuN/Cu(100). The position of the Fe adatom in the $z$-direction
is $z(\mathrm{Fe})=$\unit[-0.158]{\AA}.}
\lineup
\begin{indented}
\item[]\begin{tabular}{@{}lllllllll@{}}
\br
{\bf Fe}               & \multicolumn{4}{c}{Minority spin $\downarrow$}
               & \multicolumn{4}{c}{Majority spin $\uparrow$} \\ \ns
 & \crule{4} & \crule{4} \\
Orbital        & \multicolumn{1}{c}{$\epsilon$ (eV)}
               & \multicolumn{1}{c}{$n$}
               & \multicolumn{1}{c}{$s$ (\AA$^2$)}
               & \multicolumn{1}{c}{$z$ (\AA)}
               & \multicolumn{1}{c}{$\epsilon$ (eV)}
               & \multicolumn{1}{c}{$n$}
               & \multicolumn{1}{c}{$s$ (\AA$^2$)}
               & \multicolumn{1}{c}{$z$ (\AA)} \\
\mr
$s$            &  \01.59 & 0.22 & 2.34 & -3.13 &  \00.50 & 0.41 & 2.02 & -3.16 \\
$d_{3z^2-r^2}$ & -0.31 & 0.49 & 0.76 & -2.18 & -3.26 & 0.97 & 0.47 & -2.23 \\
$d_{zx}$       &  \00.63 & 0.10 & 0.98 & -2.12 & -3.56 & 1.00 & 0.43 & -2.30 \\
$d_{zy}$       &  \00.22 & 0.26 & 0.50 & -2.35 & -3.33 & 0.99 & 0.51 & -2.33 \\
$d_{x^2-y^2}$  & -0.30 & 0.72 & 1.29 & -2.17 & -2.65 & 0.90 & 0.70 & -2.23 \\
$d_{xy}$       &  \00.52 & 0.19 & 0.47 & -2.27 & -3.19 & 0.99 & 0.47 & -2.27 \\
$d$ total      &       & 1.76 &      &       &       & 4.85 & \\
\br
\end{tabular}
\end{indented}
\end{table}

To further substantiate our claim of the similarities between Co and
Fe, which are neighbours in the periodic system differing by a single
electron, we have performed a detailed comparative study using the
density functional theory (DFT) \cite{hohenberg1964, kohn1965}. We
used the PWSCF code (from the Quantum Espresso package,
http://www.quantum-espresso.org/, Ref.~\cite{giannozzi2009}) with a
plane-wave basis set, ultra-soft pseudopotentials, and
Perdew-Burke-Ernzerhof $\sigma$-GGA exchange-correlation functional
\cite{vanderbilt1990, perdew1996, giannozzi2009}. The kinetic-energy
cutoff was \unit[35]{Ry} and the density cut-off \unit[400]{Ry}. We
used an 8x8x4 Monkhorst-Pack mesh of $k$-points in the
self-consistent-field band-structure calculation (but 6x6x1 for the
initial relaxation calculation). The cold-smearing by \unit[0.035]{Ry}
has been performed using the Marzari-Vanderbilt scheme
\cite{marzari1999}. We used the experimentally-determined lattice
constant for Cu, $a=\unit[0.361]{nm}$. The substrate was modeled
using a 2-by-2 supercell in the lateral direction consisting of 4 Cu
layers (i.e., 37 atoms in total including the N atoms and the
adsorbate). The Cu atoms in the bottom-most layer were fixed, while
all others were allowed to relax.  We note that similar calculations
have already been performed for Fe and Mn adatoms
\cite{hirjibehedin2007} and recently also for Mn chains
\cite{rudenko2009} on the same surface.

To gain more insight into the orbital structure of the impurities, we
have also calculated the maximally localized Wannier functions (MLWF)
\cite{marzari1997, souza2001} on the impurity site using the Wannier90
package (http://www.wannier.org/, Ref.~\cite{mostofi2008}). The MLWFs
provide a very compact and accurate local representation of the
electronic structure: they give direct insight into the nature of the
chemical bonding. A 4x4x2 uniform grid of $k$-points was used here.
The initial projections consisted of $d$ and $s$ states on the adatom
and the Cu atoms, and of $sp3$ hybridized states on the N atom; in
addition, $s$ states were added in the interstitial sites of the
lattice to describe the delocalized electrons (1 per Cu atom). The
energy window was chosen so as to encompass all $d$ and $s$ levels of
the adatom.

\begin{figure}[htbp]
\centering
\includegraphics[clip,width=12cm]{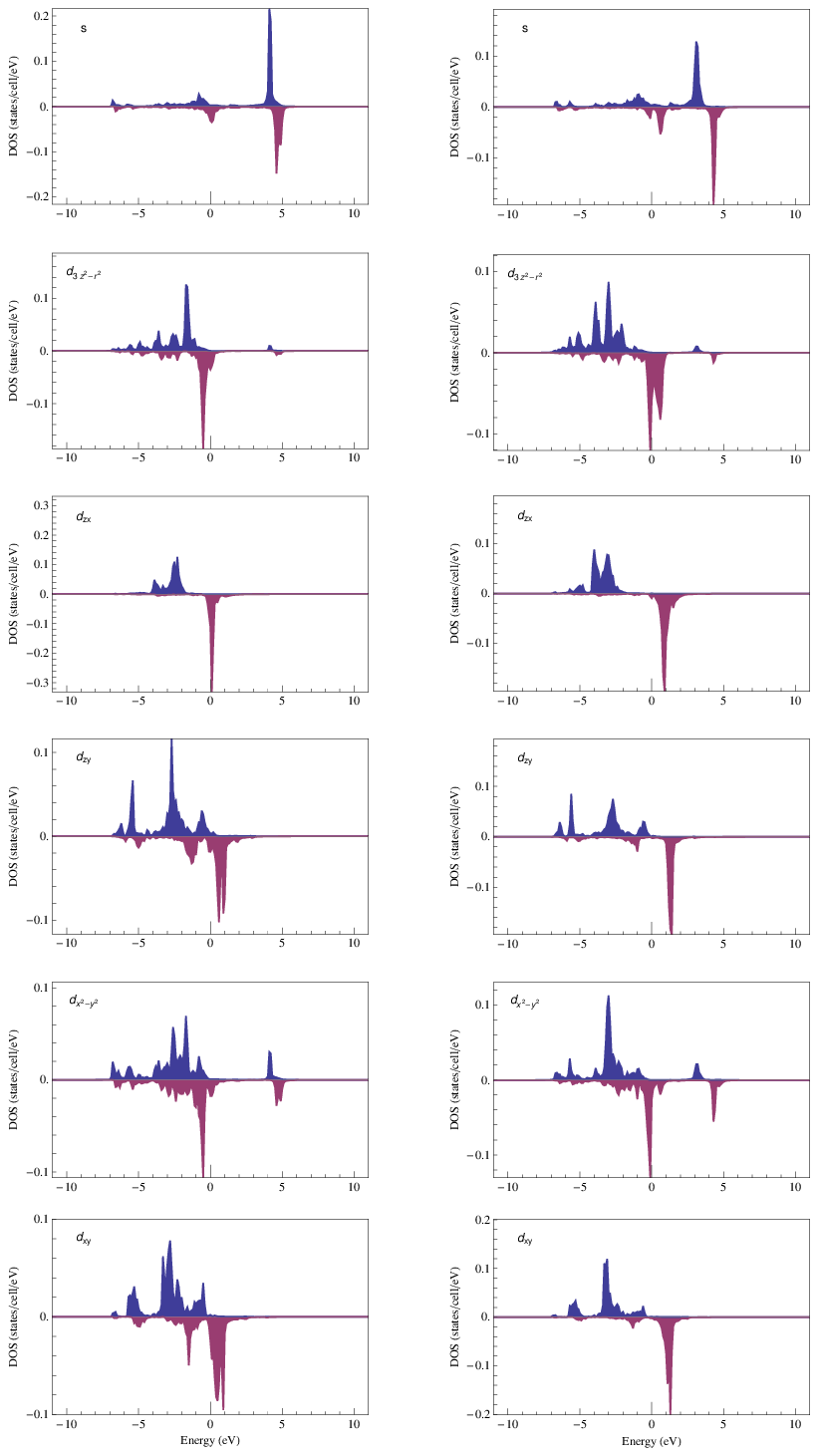}
\caption{Partial density of states (spectral functions) for the MLWF
  on Co (left) and Fe (right) adatoms. We plot both spin projections:
  ``majority'' spin (above), ``minority'' spin (below). The spectral
  functions were computed on a 16x16x3 grid of $k$-points with
  a \unit[0.1]{eV} binning resolution.}
\label{dos1}
\end{figure}

Table~\ref{table0} presents an overview of the results: the magnetic
moment and the occupancies of the adatom levels. The results are
expected -- the two adatoms differ by one electron in the $d$ orbital.
More details can be found in Tables~\ref{table1co} and \ref{table1fe}
where we show the orbital energies (defined as the average energy of
the spectral function of a particular MLWF), occupancies (defined as
the integral of the spectral function of MLWF up to the Fermi energy),
the centers of the MLWFs, as well as their spreads. It is manifest
that the difference consists of one additional electron in the
``minority'' spin orbitals for the Co adatom. In fact, the spectral
functions shown in Fig.~\ref{dos1} reveal that the ``majority'' spin
spectral functions are essentially the same in both cases, while the
``minority'' spin spectral functions are to a good approximation just
rigidly shifted.

\section{Calculation of the magnetic susceptibility in anisotropic models}
\label{appb}

In general, the magnetic susceptibility is calculated as
\cite{kubo1954, bulla2008}:
\begin{equation}
\label{chitot}
\chi_{\alpha}(T) = (g\mu_B)^2 \int_0^\beta d\tau
\left( 
\braket{S_{\alpha}(\tau) S_{\alpha}(0)}
-
\braket{S_{\alpha}}^2 
\right),
\end{equation}
where $S_{\alpha}$ is the $\alpha \in \{ x,y,z \}$ component of the
spin and $\beta=1/k_B T$. The impurity contribution to the total
magnetic susceptibility, $\chi_{\mathrm{imp},\alpha}$, is then
obtained by subtracting the susceptibility of a system without the
impurity, $\chi^{(0)}_\alpha(T)$, from the magnetic susceptibility of
the full system (i.e., the continuum of the conductance-band electrons
plus the impurity), $\chi_{\mathrm{full},\alpha}(T)$:

\begin{equation}
\chi_{\mathrm{imp},\alpha}(T) =
\chi_{\mathrm{full},\alpha}(T)-\chi^{(0)}_{\alpha}(T).
\end{equation}

In the problems with $\mathrm{U}(1)$ spin symmetry, $S_{z}$ is a
conserved quantum number and it is used to classify the invariant
subspaces of the Hilbert space, thus the calculation of the
expectation values in \eref{chitot} is particularly simple for
$\alpha=z$:
\begin{equation}
\chi_z(T) = (g\mu_B)^2 \beta \left(
\braket{S_z^2}(T) - \left[\braket{S_z}(T) \right]^2
\right),
\end{equation}
where $\braket{S_z}(T)=(1/Z^{(N)})\sum_{S_z,n} S_z \exp(-\beta
E^{(N)}_{S_z;n})$ and similarly for $\braket{S_z^2}(T)$. Here
$Z^{(N)}$ is the partition function at shell $N$, defined as $Z^{(N)}
=\Tr_{N}(e^{-\beta H}) = \sum_{S_z,n} \exp(-\beta E^{(N)}_{S_z;n})$,
$S_z$ the conserved spin quantum number, $n$ a multi-index composed of
all the remaining quantum numbers used to classify the states, and
$E^{(N)}_{S_z;n}$ the energy of the state with given quantum numbers.
For other directions $\alpha$, and for more general problems without
any symmetry in the spin space, a full dynamical calculation of the
correlator $\braket{S_{\alpha}(\tau) S_{\alpha}(0)}$ has to be
performed.

The calculation of $\chi_{\alpha}$ requires an iteration of the
operator matrix elements for all components of the total spin operator
$S_{\alpha}$. The approach is similar to the calculation of the matrix
elements of an arbitrary {\sl local} impurity operator, but it is
extended to allow for {\sl total} (``global'') spin operators
$S_{\alpha}$. The matrix elements at a given step $N+1$ are calculated
from those at the previous step by using the unitary basis
transformation resulting from the diagonalization of the Hamiltonian
matrices. We follow the notation of Ref.~\cite{bulla2008}; see
Eqs.~(39,40,43) in the cited work:
\begin{equation}
\ket{w}_{N+1} = \sum_{rs} U(w,rs) \ket{r;s}_{N+1},
\end{equation}
with
\begin{equation}
\ket{r;s}_{N+1} = \ket{r}_N \otimes \ket{s(N+1)}.
\end{equation}
Here $r$ indexes the eigenpairs of the chain Hamiltonian at the
previous iteration, i.e., $H_N \ket{r}_N = E_N(r) \ket{r}_N$, the
states $\ket{s(N+1)}$ constitute a basis for the added site, while
$U(w,rs)$ is the unitary matrix which diagonalizes the Hamiltonian
$H_{N+1}$, so that $H_{N+1} \ket{w}_{N+1} = E_{N+1} \ket{w}_{N+1}$. We
write $\hat{\vc{S}}_\alpha^{N+1} = \hat{\vc{S}}_\alpha^{N} +
\hat{\vc{s}}_\alpha^{N+1}$,
where $\hat{\vc{S}}^{N}$ is the total spin operator up to (and including)
the site $N$, while $\hat{\vc{s}}^{N+1}$ denotes the local spin operator on
site $N+1$. The matrix elements are then updated as follows [compare
with (75) in Ref.~\cite{bulla2008}]:
\begin{eqnarray}
M^{N+1}_{ww'} &= 
{}_{N+1}\bra{w} \hat{S}_{\alpha}^{N+1} \ket{w'}_{N+1} \\
&=
\sum_{rs} U(w,rs) {}_{N+1}\bra{r;s} 
\hat{S}_{\alpha}^{N+1} 
\cdot \sum_{r's'} U(w',r's') \ket{r';s'}_{N+1} \\
&= \sum_{rs,r's'} U(w,rs) U(w',r's') 
\Bigl( 
{}_{N}\bra{r} \hat{S}_{\alpha}^N \ket{r}_{N}
\delta_{ss'} \\
&\quad\quad+
\delta_{rr'}
\bra{s} \hat{s}_{\alpha}^{N+1} \ket{s'}
\Bigr) \\
&=
\label{part1}
\sum_s \sum_{rr'} U(w,rs) M^{N,s}_{rr'} U(w,r's) \\
&\quad\quad+ \label{part2}
\sum_{ss'} \left( \sum_r U(w,rs) U(w',rs') \right)
\bra{s} \hat{s}_\alpha \ket{s'}.
\end{eqnarray}
The first part of the expression on the right-hand side, \eref{part1},
is exactly the same as the conventional prescription for the recursive
calculation of the matrix elements of the operators which are singlets
with respect to the symmetry group of the Hamiltonian: for each
contribution $s$, one performs a similarity transformation of the
matrix $M^{N,s}$ from the previous iteration. The second part of the
expression, \eref{part2}, consists in adding the matrix elements of
the spin operator for the additional site weighted by the
inner-products between the columns of submatrices of the unitary
transformation $U$.

At each iteration, the required expectation values are calculated in the
same way as for local singlet operators [cf. (58) in
Ref.~\cite{bulla2008}]:
\begin{equation}
\braket{S_{\alpha}}(T_N) \approx
\frac{1}{Z^{(N)}}
\sum_{w} M^N_{ww}
\end{equation}
where the matrix elements $M^N_{ww}$ are those that correspond to the
operator $S_{\alpha}$.

The imaginary-time integral of the correlator
$C(\tau)=\braket{S_{\alpha}(\tau) S_{\alpha}(0)}$ may be expressed
using the Lehmann representation as
\begin{equation}
\int_0^\beta C(\tau) \dr\tau = \frac{1}{Z^{(N)}}
\sum_{w,w'} 
\left(M^N_{ww'} \right)^*
M_{w'w}^N 
\frac{e^{-\beta E_w} - e^{-\beta E_{w'}}}
{\beta E_{w'} - \beta E_{w}}.
\end{equation}
Note that for the spin operator along the $y$ direction, the NRG
calculation has to be performed using complex numbers.

Where comparisons can be made (e.g., in isotropic models) the magnetic
susceptibility calculated in the traditional way and using the
proposed technique differ by no more than a few per mil in calculations
performed for $\Lambda=2$, and even less for higher $\Lambda$. We
observed, however, that more states need to be kept in this approach
than in the standard NRG thermodynamic calculations. The calculation of
$\chi$ is made possible by the property of the energy-scale separation
of total spin operators, much like the entire NRG iteration is based
on the energy-scale separation property of the Hamiltonian.

The approach described in this section can be immediately generalized to the
calculation of the out-of-diagonal components of the magnetic susceptibility
tensor.

\ack
RZ acknowledges the support of the Slovenian Research Agency (ARRS)
under grant no. Z1-2058.
TP acknowledges financial support by the DFG through SFB 602.

\bibliography{fe}

\end{document}